\documentclass[runningheads]{llncs}

\usepackage{xcolor}
\usepackage{amsmath}
\usepackage{graphicx}
\usepackage{amssymb}
\usepackage{graphicx}
\usepackage{booktabs} 
\usepackage{graphicx}
\usepackage[ruled,vlined]{algorithm2e}

\begin{document}
\title{Explainability Guided Multi-Site\\ COVID-19 CT Classification}
\titlerunning{Explainability Guided Multi-Site COVID-19 CT Classification}
%
\author{Ameen Ali \and
Tal Shaharabany \and
Lior Wolf}
\authorrunning{Ameen Ali et al.}
%
\institute{School of Computer Science, Tel Aviv University, Tel Aviv, Israel \\
}

\maketitle              
\begin{abstract}
Radiologist examination of chest CT is an effective way for screening COVID-19 cases. In this work, we overcome three challenges in the automation of this process: (i) the limited number of supervised positive cases, (ii) the lack of region-based supervision, and (iii) the variability across acquisition sites. These challenges are met by incorporating a recent augmentation solution called SnapMix, by a new patch embedding technique, and by performing a test-time stability analysis. The three techniques are complementary and are all based on utilizing the heatmaps produced by the Class Activation Mapping (CAM) explainability method. Compared to the current state of the art, we obtain an increase of five percent in the F1 score on a site with a relatively high number of cases, and a gap twice as large for a site with much fewer training images. 
\end{abstract}

\section{Introduction}

Deep neural networks are currently the leading image classification method. Their ability to generalize is well-documented. However, in many medical imaging domains, one faces challenges that reduce the effectiveness of the generic solutions. First, due to the cost of acquisition, privacy issues, and the expertise required for labeling, the typical datasets are smaller than those available for many other computer vision tasks. Second, in medical images, the exact capturing apparatus, its setting and its operators all can greatly affect the distribution of the obtained images, causing a sizable domain shift. Third, many diseases are manifested through symptoms that are well localized in images, while the supervision is given at the image-level.

In this work, we demonstrate that explainability methods, which link the classification outcome to specific image regions, can provide an important building block for overcoming the three issues. First, the heatmap obtained from such methods serves as the basis of an augmentation method called SnapMIX~\cite{huang2021snapmix}, which we demonstrate is also effective for the COVID-19 classification task we study in this work. Second, the heatmap can provide a delineation of whether or not local image patches are strongly linked to the obtained classification. By requiring that image patches of similar relevancy have similar embedding, we can improve the classification performance. Third, we can use the heatmap in order to validate, at test time, the stability of the obtained classification by perturbing the image locations that are the most relevant to the prediction. If the majority of the perturbations do not support the prediction, we flip the predicted label.

We evaluate our method with well-established benchmarks for the classification of Computed Tomography (CT) scans as COVID-19 positive or COVID-19 negative, and present clear evidence for the utility of our method. The gap in performance we obtain is larger than the variance between the state of the art methods. On site A, in which performance (F1 score and accuracy) is over 90\%, we improve to over 95\%. On site B, in which the performance levels are in the high 70 percentages, we obtain results of almost 90\%. 

\section{Related Work}

\noindent{\bf COVID19 Classification\quad} The SARS COV-2 infection (COVID-19) has  a devastating impact on the respiratory system and has caused an enormous number of deaths. In the last year, many deep learning methods were developed for classifying COVID-19 in 2D or 3D medical images~\cite{gozes2020rapid,rahimzadeh2020modified,zhang2020clinically,wang2020covid}. Some recent methods use transfer learning from  models pretrained on ImageNet~\cite{hall2020finding,apostolopoulos2020extracting}. 

Following Wang et al.~\cite{Wang_2020}, we study classification in two CT datasets. To overcome the domain shift, their approach adds a contrastive loss that decreases the differences between the latent space distributions. 
Unlike previous work in the domain of CT diagnosis of COVID-19, our method employs a generic ResNet architecture and our contribution lies solely in the training procedure and in the inference procedure.

\noindent{\bf Data augmentation\quad} Many augmentation approaches were developed over the years as a form of regularization. These include geometric transformations~\cite{taylor2017improving} and color space transformations~\cite{wu2015deep}, which have been shown to improve many medical applications~\cite{litjens2017survey}. 

Data mixing approaches create virtual samples that combine multiple images from different categories. The generated image has a fuzzy label from the two categories. In MixUp~\cite{guo2019mixup}, the augmented image is a linear interpolation between two different images. The fuzzy labels are computed using the same  weights as the images. Cutmix~\cite{yun2019cutmix} extracts a box from one image and pastes it to the second. The fuzzy labels are proportional to the area of the box. SnapMix~\cite{huang2021snapmix} is similar to Cutmix, except that the area of the patch is replaced by the sum of the CAM activations within the extracted and the masked patches. It was shown to be highly effective on find-grained classification datasets of natural images. Here, it is applied to the binary classification of medical images.

\noindent{\bf Explainability\quad} The task of generating a heatmap that indicates local relevancy from the perspective of a CNN observing an input image has been tackled from many different directions, including gradient-based methods~\cite{shrikumar2017learning,srinivas2019full,selvaraju2017grad}, attribution methods~\cite{bach2015pixel,montavon2017explaining,nam2019relative,gur2020visualization}, and image manipulation methods~\cite{fong2019understanding,fong2017interpretable,lundberg2017unified}.

The CAM method~\cite{zhou2016learning} is based on the gradient of the loss with respect to the input of each layer. CAM and its extension GradCAM~\cite{selvaraju2017grad} have been used by downstream applications, such as weakly-supervised semantic segmentation~\cite{li2018tell}. Here, we make a novel use of CAM for creating more effective patch embeddings and for test time augmentation.

\noindent{\bf Contrastive learning\quad} The loss we employ between patches of different levels of relevancy is related to contrastive learning methods that have recently made a large impact in the field of self-supervised learning, where it is often used to link an image to its transformed version~\cite{he2019momentum,misra2019self,chen2020simple}. Our work is applied at the patch level.  Contrastive learning has emerged in metric learning~\cite{chopra2005learning} and subsequently in unsupervised representation learning~\cite{hadsell2006dimensionality}. The learned embedding brings closer associated samples, while distancing other samples. In our case, association is being determined by the CAM-derived relevancy.

\section{Method}

Our experiments utilize a Resnet-50 network~\cite{he2016deep}, trained with the conventional binary cross entropy loss $L_{\text{BCE}}$, as the baseline classifier. We then apply (i) SnapMix~\cite{huang2021snapmix}, (ii) a novel optimization term called the Contrastive Patch Embedding loss, and (iii) a novel test time voting procedure. All three techniques utilize the heatmaps produced by the CAM method~\cite{zhou2016learning}.

\subsection{SnapMix~\cite{huang2018weakly}}

\begin{figure}[t]
\begin{tabular}{ccccc}
\includegraphics[width=0.198\linewidth]{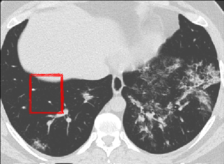} & 
\includegraphics[width=0.198\linewidth]{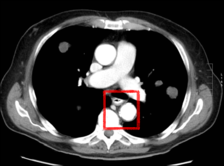} &
\includegraphics[width=0.198\linewidth]{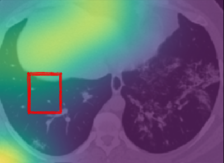} &
\includegraphics[width=0.198\linewidth]{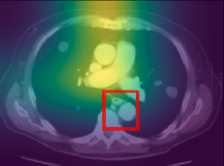} &
\includegraphics[width=0.198\linewidth]{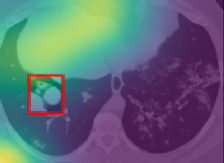} \\
(a) & (b) & (c) & (d) & (e)\\
\end{tabular}
\caption{An illustration of the SnapMix process. (a) A first random image, in this case a positive image form site A, (b) a second random image, shown is a negative image from site B. A random box is marked in each image. (c,d) the CAM maps of (a,b) respectively, with the associated boxes marked. (e) The SnapMix training image obtained by combining images (a) and (b) based on the random boxes. The label of the virtual training image is determined by mixing the labels of the two source images in accordance with the sum of the CAM activations in each box.}
\label{fig:snapmix}
\end{figure}

The SnapMix method is illustrated in Fig.~\ref{fig:snapmix}. It combines two training images, depicted in panels (a) and (b) by considering a random box in each image (marked in red). The importance of each of the two boxes is evaluated by integrating the CAM scores in each box (panels c,d). The virtual sample is generated by pasting the box from the second image onto the selected box of the first image (panel e), and labeling the new image proportionally to the integrated CAM scores. More specifically, a ratio ($\rho_a$,$\rho_b$) is computed for each image by considering the sum of all CAM scores in a box over the sum of the CAM scores of the entire image. The labels are then linearly interpolated between the labels of the two images, using the the complement of the obtained box ratio in the first image ($1-\rho_a$) and the ratio in the second image $\rho_b$. 

Unlike the original experiments in~\cite{huang2021snapmix}, which considered datasets
with many classes, in our case the problem is binary. It often happens that both images are of the same class. Moreover, since we train using images from two sites, the virtual images created can potentially play a role in overcoming the domain shift.

\subsection{Contrastive Patch Embedding}
\label{sec:cpe}

The input images we receive are of size $224\times 224$, the receptive field of the ResNet-50 architecture is of size $32\time 32$, and the embedding is of spatial dimensions of $7 \times 7$ with a depth of $2,048$. For each of the $7\times 7=49$ vectors in $\mathbb{R}^{2048}$, we compute the sum of the CAM activations in the associated patch of size $32\times 32$. We then select four vectors out of the 49: two with the highest sum of activations $u_1$ and $u_2$, and two with the lowest sum $v_1$ and $v_2$.

The embedding loss we propose is a contrastive loss~\cite{wu2018unsupervised,he2019momentum,oord2018representation} that considers the dot products between the four vectors. 
\begin{align}
L_\text{CPE}(u_1,u_2,v_1,v_2) =& -\ln{ 
\frac{\exp (u_1^\top u_2)}{\exp (u_1^\top u_2) + \sum_{i,j=1}^2 \exp (u_i^\top v_j)}  }\nonumber \\ & -\ln{ 
\frac{\exp (v_1^\top v_2)}{\exp (v_1^\top v_2) + \sum_{i,j=1}^2 \exp (u_i^\top v_j) }  }.
\label{eq:cpe}
\end{align}

This loss brings together the two most label-supporting embedding-vectors and two most label-opposing embedding-vectors. At the same time, it distances the top label-supporting embedding-vectors form the pair of vectors that support the alternative label.

\begin{figure}[t]
    \centering
    \begin{tabular}{cccc}
             \includegraphics[width=0.24\linewidth]{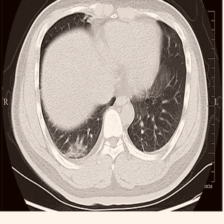} &
             \includegraphics[width=0.24\linewidth]{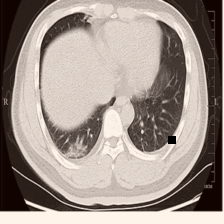} & 
             \includegraphics[width=0.24\linewidth]{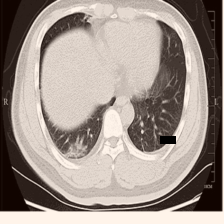} & 
             \includegraphics[width=0.24\linewidth]{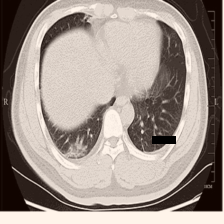} \\
             (Original, 0.47) & (Image 1, 0.61) & (Image 2, 0.59) & (Image 3, 0.64)\\
                          \includegraphics[width=0.24\linewidth]{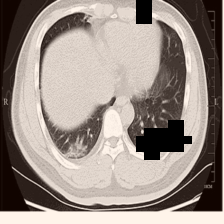} & 
              \includegraphics[width=0.24\linewidth]{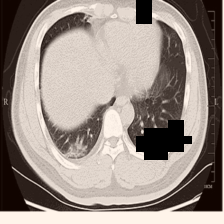} & 
             \includegraphics[width=0.24\linewidth]{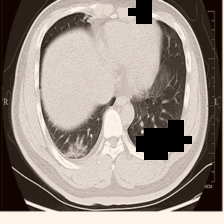} & 
             \includegraphics[width=0.24\linewidth]{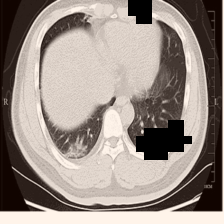} \\
             (Image 28, 0.90) & (Image 29, 0.90) & (Image 30, 0.93) & (Image 31, 0.92)\\
    \end{tabular}
    \caption{The original test time image (top left) got a negative classification score, with a probability of 0.47 of being positive. Removing even a small number of patches (Images 1-3) increased this probability to be above 0.5.  Subsequently, as more and more patches were removed, the probability of a positive case increased further, and became higher than $1-\theta$, see the last derived images (out of $k=31$ images), Images 28-31.}
    \label{fig:masking}
\end{figure}

\subsection{CAM-Directed Test Time Augmentation}
\label{sec:testtime}

It may be the case that the decision for a certain label is based on local artifacts that bias the network into giving the wrong prediction. In order to avoid such cases, we classify each image $k+1$ times: using the entire image, and when masking one out of $k$ different patches.

For this purpose, we divide the image into small non-overlapping patches of size $8\times 8$, obtaining a grid of size $28\times 28$. For each cell in the grid, we compute the sum of the CAM activations. We then create $k=31$ alternative images, by masking out sequentially the $k$ patches with the highest sum of activations. In the first alternative image, we mask out the patch with the highest CAM scores; in the second, we mask out the two patches with the highest CAM scores; and so on. See Fig.~\ref{fig:masking} for an illustration.

The label that we report is obtained by performing voting among the classifier output of the $k$ images. A supporting vote occurs when the pseudo-probability obtained from the network classifier is at least $\theta=0.2$ if the original image has a positive label (i.e., a pseudo-probability larger than 0.5), or lower than $1-\theta$ for images with negative samples. If more than half the $k$ votes are not supporting, we flip the label. In other words, if the inferred labels of more than half out of the $k$ alternative images contradict, with a high certainty, the label the classifier assigns to the entire image, we flip the predicted label of the image.


\begin{table*}[t]
\caption{COVID-19 classification results for site A}
\centering
\resizebox{\textwidth}{!}
{
\begin{tabular}{@{}l@{~~}c@{~~}c@{~~~}c@{~~~}c@{}}
\toprule
Method & Accuracy & Precision & Recall & F1  \\
\midrule
  Series Adapter~\cite{rebuffi2017learning} & 85.73±0.71 & 90.98±0.79 & 81.91±2.61&86.19±1.65 \\
  Parallel Adapter~\cite{rebuffi2018efficient} & 82.13±1.91 & 83.51±1.87 & 80.02±2.47 & 82.39±1.78 \\
  MS-Net~\cite{liu2020ms} & 87.98±1.31 & 93.78±2.76 & 84.91±2.83 & 88.73±1.20 \\
  \hline
  Single (Covidnet)~\cite{wang2020covid} & 77.12±0.98 & 80.04±2.87 & 70.97±2.37 & 76.03±1.13  \\
  Single (Redesign)~\cite{Wang_2020} &89.09±1.08 &94.58±2.07 & 83.78±0.62 & 88.97±0.91 \\
  Joint (Covidnet)~\cite{wang2020covid} & 68.72±1.94 & 68.27±1.21 & 69.41±3.91 & 69.17±1.93 \\
  Joint (Redesign)~\cite{Wang_2020} & 78.42±2.19 &80.82±1.05&74.07±3.16&77.86±2.01 \\
  \hline
SepNorm~\cite{Wang_2020}  & 88.76±0.78 & 95.46±0.74 & 82.97±1.66 & 87.88±0.81  \\
  SepNorm + Contrastive~\cite{Wang_2020} & 90.83±0.93 & 95.75±0.43 & 85.89±1.05 &  90.87±1.29 \\
  \hline
  Baseline architecture &89.68±0.46&95.02±0.40&83.99±0.51&89.13±0.47 \\
Baseline  + CPE loss (ablation) & 91.71±1.21&97.02±1.65&85.13±1.34&90.03±0.61\\
  SnapMix~\cite{huang2021snapmix} & 92.38±0.32 & 98.33±1.81&86.42±1.50&91.92±0.37\\
SnapMix + Contrastive & 91.99±0.13 & \textbf{99.02±0.52} 
&84.22±1.22&91.03±0.83\\
  SnapMix + CPE (ablation) & 95.73±0.07 & 98.97±0.33 &92.49±0.47&95.59±0.12\\
  Our full method & \textbf{95.90±0.24} &98.64±0.12&\textbf{92.93±0.40}&\textbf{95.87±0.25}\\
\bottomrule
\end{tabular}}
\label{tab:main_resultsa}
\end{table*}

\begin{table*}[t]
\caption{COVID-19 classification results for site B}

\resizebox{\textwidth}{!}
{
\begin{tabular}{@{}l@{~~}c@{~~}c@{~~~}c@{~~~}c@{}}
\toprule

Method & Accuracy & Precision & Recall & F1 \\
\midrule
  Series Adapter~\cite{rebuffi2017learning} &70.01±3.82&63.04±4.87&74.91±1.89&67.08±3.09 \\
  Parallel Adapter~\cite{rebuffi2018efficient}  & 74.93±1.83 & 79.84±1.75 & 71.81±2.47 & 73.46±1.68\\
  MS-Net~\cite{liu2020ms} & 76.23±1.81 & 79.29±1.48 & 74.07±1.29 & 76.54±1.73\\
  \hline
    Single (Covidnet)~\cite{wang2020covid}  & 63.12±2.09 & 64.03±3.91 & 57.73±2.94 &61.09±1.28 \\
  Single (Redesign)~\cite{Wang_2020} & 77.07±1.92 &79.48±0.96&74.69±3.91&77.04±2.17\\
  Joint (Covidnet)~\cite{wang2020covid} & 63.27±2.82 & 64.27±3.81 & 54.19±4.17 & 59.78±3.12\\
  Joint (Redesign)~\cite{Wang_2020}&69.67±0.92 & 64.98±3.17 & 66.94±5.86 & 66.89±4.91\\
  SepNorm~\cite{Wang_2020}   & 76.89±0.65 & 80.74±2.98& 70.34±3.76 & 75.02±1.14 \\
  SepNorm + Contrastive~\cite{Wang_2020} & 78.69±1.54 & 78.02±1.34 & 79.71±1.42 & 78.83±1.43 \\
  \hline
  Baseline architecture & 85.23±0.41&86.54±0.84&83.58±0.81&84.51±0.62\\
Baseline  + CPE loss (ablation)& 85.96±1.22& 87.03±1.22 &84.71±1.02 &85.22±0.79\\
SnapMix~\cite{huang2021snapmix} &87.56±0.41&\textbf{88.76±0.53}&85.19±1.02&86.85±0.48\\
SnapMix + Contrastive & 87.03±0.35&88.33±0.85&84.22±0.79&85.72±0.65\\
  SnapMix + CPE (ablation) & 87.02±0.49& 88.32±0.69&84.69±1.02 &86.95±0.76\\
  Our full method & \textbf{88.76±0.26}&87.44±0.42&\textbf{88.48±0.19}&\textbf{88.25±0.22}\\

\bottomrule
\end{tabular}}
\label{tab:main_resultsb}
\end{table*}
\begin{table*}[t]
\caption{{Classification results on COVIDx-CT dataset, which is considered the biggest dataset }}
\centering
\begin{tabular}{l@{~~}c@{~~}c@{~~~}c@{~~~}c@{~~~}c@{~~~}c@{~~~}c@{~~~}c}
\toprule
&&\multicolumn{2}{c}{Sensitivity}&\multicolumn{2}{c}{PPV}&
\\
\cmidrule(lr){3-4}
\cmidrule(lr){5-6}
\
Method & Acc  & Non-Covid-19 & Covid-19 & Non-Covid-19 & Covid-19 & \\
\midrule 
ResNet-50 \cite{he2016identity}&98.7\%&98.7\%&96.2\%&97.8\%&99.1\% \\
NASNet-A-Mobile\cite{zoph2018learning}&98.6\%&97.9\%&96.8\%&99.6\%&97.1\% \\
EfficeintNet-B0 \cite{tan2019efficientnet} &98.3\%&97.8\%&95.8\%&98.7\%&98.6\% \\
COVIDNet-CT \cite{gunraj2020covidnet}&99.1\%&99.0\%&97.3\%&98.4\%&99.7\% \\
Ours &\textbf{99.5\%}&\textbf{99.7\%}&\textbf{99.7\%}&\textbf{99.8\%}&\textbf{99.8\%} \\

\bottomrule
\end{tabular}
\label{tab:main_results}
\end{table*}

\begin{figure}[t]
\begin{center}
\begin{tabular}{cc}
    \includegraphics[width=.45\linewidth]{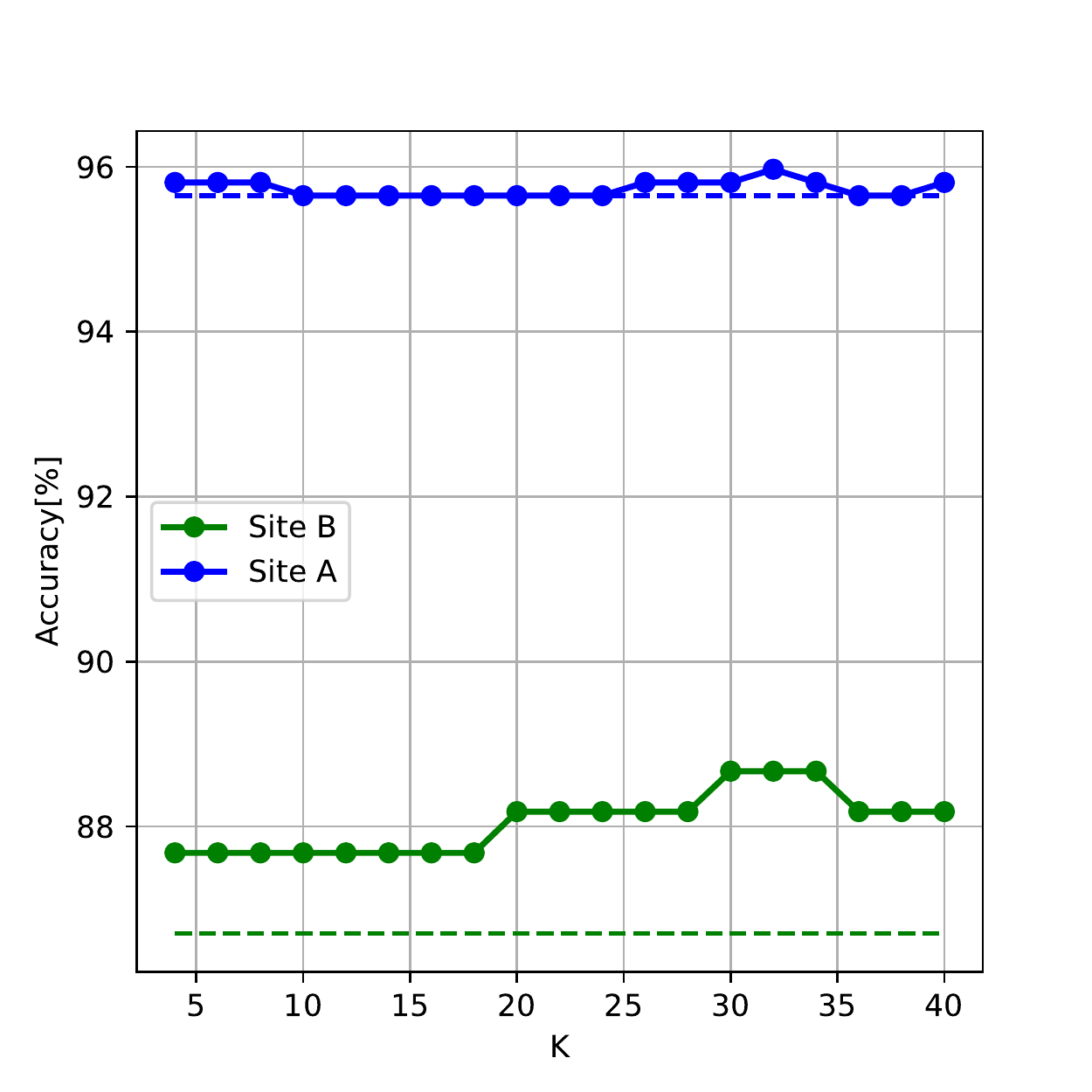}&
    \includegraphics[width=.45\linewidth]{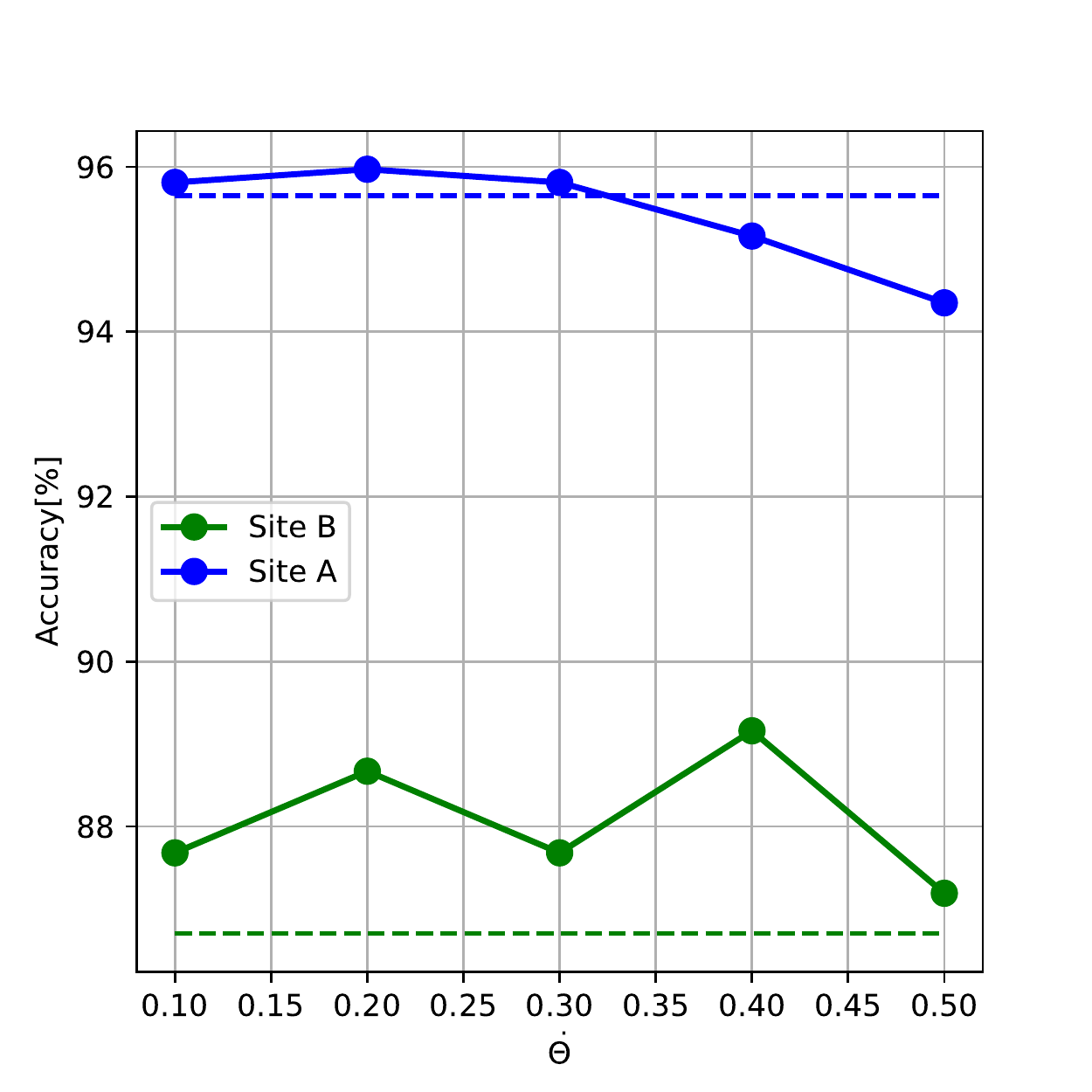}\\
    (a) & (b)\\
    \vspace{-.6cm}
\end{tabular}
\end{center}
    \caption{Parameter sensitivity analysis of the test time augmentation method. (a) The effect of varying the number of alternative images $k$, (b) The effect of varying the the certainty probability threshold $\theta$.}
    \label{fig:ablation}
\end{figure}

\section{Experiments}

\noindent{\bf Data\quad} We evaluate the proposed method on three datasets, the first two contains CT images of patients who are COVID-19 positive or negative.  The SARS-CoV-2 dataset (denoted as site A) consists of 2,482 CT images from 120 patients, in
which 1252 are positive with COVID-19. The 1,230 negative samples are inflicted with other lung disease. The resolution of these images vary between $119 \times 104$ and $416 \times 512$. The COVID-CT dataset~\cite{zhao2020covid} (denoted as site B) is much smaller and includes 349 CT images from 216 COVID-19 positive patients and 397 CT images from 171 control patients. The resolution of the images of site B ranges from $102\times 137$ to $1853 \times 1485$. Following~\cite{Wang_2020}, the images of both datasets are resized to a fixed resolution of $224 \times  224$ and are
intensity normalized to zero mean and unit variance. The classification accuracy,  F1 score, Sensitivity, and Precision are reported in percents using the train/test splits of the different datasets.

The third dataset we employ is COVIDx-CT \cite{gunraj2020covidnet} this dataset is considered one of the largest in terms of the number of annotated samples provided, containing 35996 training images of negative samples and 82286 of positive samples, the test split contains 12245 and 6018 samples for positive and negative patients respectively. For the quantitative analysis we report accuray as well as sensitivity and PPV (positive predictive value) for each infection type at the image level.

\noindent{\bf Implementation Details\quad} The architecture of our model is based on ResNet50, followed by an MLP classifier, the ResNet model is initialized  with pretrained ImageNet weights. We train the model for 200 epochs. The cross entropy loss is used unweighted on the original samples or on virtual SnapMix samples, as dictated by a beta distribution with a parameter of $\alpha=1$, which is the default parameter in~\cite{huang2021snapmix}. The $L_\text{CPE}$ loss is applied to all samples and is summed, unweighted with the cross entropy loss.

\noindent{\bf Baseline methods\quad} The first two baseline methods used for sites A and B are methods that address domain shift in medical images. Series Adapter~\cite{rebuffi2017learning} and Parallel Adapter~\cite{rebuffi2018efficient} include a domain adapter model that is based on a filter bank, in order to learn a joint representation from multiple datasets. MS-Net~\cite{liu2020ms} was originally developed for a multi-site prostate segmentation task. It uses domain-specific auxiliary decoders. For classification tasks, each site is associated with an auxiliary classification head. The results for these three methods are obtained from~\cite{Wang_2020}.

The single and joint methods from~\cite{wang2020covid}, employ an architecture called Covidnet. The difference is whether the method is trained on each dataset separately or not. It was also rerun in~\cite{Wang_2020}, using a modified architecture (redesign). The SepNorm method of~\cite{Wang_2020} uses features that are normalization for each site separately. It is further augmented with a contrastive loss that minimizes the domain shift (``SepNorm + Contrastive'').

We present results for the ResNet-50 based architecture that our method utilizes (``Baseline architecture''), and also study the effect of our CPE loss (Eq.~\ref{eq:cpe}) on it (``Baseline+CPE loss''). Results are also presented when augmenting this architecture with the SnapMix method. As additional ablations, we present result for SnapMIX combined with either the contrastive loss of~\cite{Wang_2020} (``SnapMix+Contrastive loss'') or with our CPE loss (``SnapMix + CPE''). Finally, we present our full method, which includes SnapMix augmentation, the CPE loss, as well as the CAM-driven test time augmentation and voting. For the COVIDx-CT dataset we compare our method with the reported baselines in \cite{gunraj2020covidnet},  the COVIDNet-CT baseline \cite{gunraj2020covidnet} was pre-trained on ImageNet \cite{deng2009imagenet} and later finetuned on COVIDx-CT \cite{gunraj2020covidnet} dataset using stochastic gradient descent with momentum \cite{qian1999momentum}. We also compare our model with existing models for image recognition (ResNet50 , EfficeintNet-B0 , NASNet-A-Mobile \cite{he2016identity,zoph2018learning,tan2019efficientnet}) for image recognition finetuned on COVIDx-CT dataset.

\noindent{\bf The results} are reported in Tab.~\ref{tab:main_resultsa} for site A, and Tab.~\ref{tab:main_resultsb} for site B. Evidently, for both sites, the baseline architecture is already competitive with the best method from the literature, which is SepNorm with the Contrastive loss. In site A the baseline is slightly inferior, and for site B it is considerably preferable. 

Adding the CPE loss (Sec.~\ref{sec:cpe}) improves the results on both sites. So does the SnapMix augmentation, by a larger amount. The two contributions are complementary and adding both the CPE loss and SnapMix provides considerably better results than both separately in site A. In site B, the combination of both provides a slightly higher F1 score than either contributions alone. However, SnapMix by itself is slightly better in terms of the three other scores. The ablation of using the contrastive loss of~\cite{Wang_2020} combined with the SnapMix technique hurts the F1 performance relative to SnapMix in site A, by increasing the precision on the expense of the recall. On site B, it hurts all four scores. 

Our complete method, which adds the test time augmentation of Sec.~\ref{sec:testtime} on top of SnapMix and the CPE loss, obtains the best accuracy, recall, and F1 score among all methods. Its precision is slightly less than the best ablation method. However, the gap in performance in the F1 score (which combines both precision and recall) is substantial in comparison to the ablation method with the highest precision (almost 5\% in site A, and 1.5\% in site B). In Tab~\ref{tab:main_results} we show the results for the COVIDx-CT dataset, as shown our full model achieves superior performance over all of the reported baselines.

\noindent{\bf Parameter Sensitivity\quad} SnapMix employs the default augmentation parameters prescribed by~\cite{huang2021snapmix}. The CPE loss is defined without a temperature parameter that is commonly used in other contrastive learning methods and it employs the minimal number of patches. It is, therefore, virtually parameter-free. 

The parameter sensitivity of the CAM-driven test-time voting is explored in Fig.~\ref{fig:ablation}, in which performance without this voting (``SnapMix+CPE'') is depicted as a dashed horizontal line. When varying the number of augmented images $k$ (panel a), we observe that for any value of $k$, there is a performance boost for site B, and this is maximized between $k=30$ and $k=35$. The performance boost for site $A$ is smaller for all $k$, and peaks at the value of $k=31$. However, no value of $k$ hurts the performance in site $A$. 

Varying the value of the probability threshold depicted in Fig.~\ref{fig:ablation}(b), shows that there is a positive benefit for all tested values $\theta\in \{0.1,0.2,0.3,0.4,0.5\}$ considering site B. The largest contribution is for the value of 0.4. For site $A$, however, the contribution is positive only for conservative values (smaller than 0.3, when flipping the label of the test image becomes less frequent). The value of $\theta=0.2$ provides a small boost to site A and is also the 2nd highest for site B.

\section{Conclusions}

We present a method of COVID-19 detection in CT scans. The method tackles many of the challenges faced by medical imaging classification systems: distribution shifts across sites, limited training data, and the lack of region based tagging. We propose to combine three different techniques, which have in common the reliance on the heatmap produced by the CAM explainability method. The first method is a powerful regularizer called SnapMix, which was previously used for fine-grained classification. The second is a novel patch embedding method that considers the two patches that show the strongest CAM activations in a given image and the two that present the lowest activations. Finally, we propose a voting method that constructs multiple masked images based on the CAM score. Taken together, our method obtains, despite using a generic network architecture, state of the art results on the two publicly available COVID-19 CT datasets. The gap in performance is extremely sizable, and we demonstrate the individual contribution of each component to it.

\section*{Acknowledgment}
This project has received funding from the European Research Council (ERC) under the European Unions Horizon 2020 research and innovation programme (grant ERC CoG 725974).

\bibliography{mixup}
\bibliographystyle{splncs04}
\end{document}